\documentclass{ws-ijmpa}
\def\1{\mbox{l\hspace{-0.53em}1}}
\usepackage{amsfonts}
\DeclareMathAlphabet{\mathbbm}{U}{bbm}{m}{n}
\SetMathAlphabet\mathbbm{bold}{U}{bbm}{bx}{n}
\begin{document}
\title{MASSES OF $[{\bf 70},\ell^+]$ BARYONS IN LARGE $N_c$ QCD}
%Proceedings\footnote{\uppercase{T}his work is supported by etc, etc.}}

\author{N. Matagne and Fl. Stancu}
%\footnote{\uppercase{W}ork partially
%supported by grant 2-4570.5 of the \uppercase{S}wiss
%\uppercase{N}ational \uppercase{S}cience \uppercase{F}oundation.}}

\address{University of Li\`ege, Physics Department,\\
Institute of Physics, B.5, \\
Sart Tilman, B-4000 Li\`ege 1, Belgium\\
E-mail: nmatagne@ulg.ac.be,
fstancu@ulg.ac.be}

\maketitle

\begin{abstract}
Previous work is extended from SU(2) to SU(3) and we present results for the mass spectrum of the $[{\bf 70},\ell^+]$ ($\ell=0,2$) nonstrange and strange baryons in the $1/N_c$ expansion. We show that the dominant term is the spin-spin interaction and its contribution vanishes at large excitations.  
\end{abstract}

\section{Introduction}
In the low energy region, typical for baryon spectroscopy, QCD does not admit a classical perturbative expansion because the coupling constant is too large. Another kind of perturbative expansion shaped out from 't Hooft's proposal of 1974, to generalize QCD to $N_c$ colors. In this case $1/N_c$ can be used as an expansion parameter. Five years later, Witten\cite{WITTEN} analyzed properties of baryons in the large $N_c$ limit and determined power counting rules. In 1984, Gervais and Sakita\cite{gervais} and independently, in 1993,  Dashen and Manohar\cite{DM93} realized that if $N_c \to \infty$ the ground state baryons satisfy a contracted SU$(2N_f)_c$ spin-flavor algebra where $N_f$ is the number of flavors. This algebra is identical to SU($2N_f$) in the large $N_c$ limit.  For excited baryons, this symmetry is broken at the first order\cite{Goi97} in $1/N_c$. Furthermore, excited states are resonances and have widths of order $N_c^0$ [\refcite{cohen1}]. Nevertheless, the $1/N_c$ expansion has been used during the last ten  years  to describe successfully states belonging to various SU(6) excited multiplets\cite{CGKM}\cdash\cite{MS3}. Most of these studies ignore the finite width and treat the resonances as bound states.

In this paper, we are summarizing our results, presented in detail in Ref. [\refcite{MS3}], which is an extension of the previous study of the $[{\bf 70}, \ell^+]$ ($\ell=0,2$) multiplet in the $1/N_c$ expansion from $N_f=2$ (Ref. [\refcite{MS2}]) to $N_f=3$.

\section{The mass operator}
Large $N_c$ baryons belonging to the $[{\bf 70},\ell^+]$ multiplet are composed of one or two excited quarks and $\mathcal{O}(N_c)$  quarks left in the ground state. The general procedure for calculating the mass spectrum is to split the wave function into a symmetric core composed to $N_c-1$ quark and an excited quark. With such an approach, one can treat the core in the same way as the ground state.

The mass operator must be rotationally invariant, parity and time reversal even. The isospin breaking is neglected.
Then, the general $1/N_c$ expansion for the $[{\bf 70},\ell^+]$ mass operator reads
\begin{equation}
 M_{[{\bf 70},\ell^+]}=\sum^6_{i=1}c_iO_i +d_1B_1+d_2B_2+d_4B_4.
\end{equation}
where $O_i$  are rotational invariants and SU(3)-flavor scalars and the operators $B_i$ provide SU(3) breaking and  are defined to have non-vanishing matrix elements for strange baryons only.
 One has both core generators $\ell^i_c$, $S^i_c$, $T^a_c$ and $G^{ia}_c$ and excited quark generators $\ell^i_q$, $s^i$, $t^a$ and $g^{ia}$. The values of the  coefficients $c_i$ and  $d_i$ are obtained by a numerical fit to  data.

 Due to a lack of experimental data we had to make a selection among all possible operators. With the help of previous experience\cite{CCGL,SGS,MS2}, we kept only the most dominant operators in the mass formula. Table \ref{operators} shows the list of operators chosen for this study. In this table, $O_1$ is the SU(6) scalar operator linear in $N_c$. $O_2$ and $O_5$ are the dominant part of the spin-orbit and spin-spin operators respectively. The first, which acts only on the excited quark, is of order $N_c^0$ but the two-body spin-spin operator is of order $N_c^{-1}$. The operators $O_3$ and $O_4$ are of order $N_c^0$ due to the presence of the SU(6) generator $G_c^{ia}$ which sums coherently. $O_6$ represents the isospin-isospin operator, having matrix elements of order $N_c^0$ due to the presence of $T_c^a$ which sums coherently too.
 
 As already mentioned, the operators $B_i$ break the SU(3)-flavor symmetry. The operators $B_1$, $B_2$ are the standard breaking operators while $B_4$ is directly related to the spin-orbit splitting. They break the SU(3)-flavor symmetry to first order. 

The calculation of the matrix elements of these operators is not easy. One can quite directly obtain the matrix elements of all the operators but for $G_c^{ia}$. A generalized Wigner-Eckart theorem can be applied to obtain the $G_c^{ia}$ matrix elements in terms of SU(6) isoscalar factors. We have derived analytic expressions of these isoscalar factors for SU(6) symmetric wave functions\cite{MS4}. 
%\begin{center}
\begin{table}[h!]
\tbl{List of operators and the coefficients resulting from the fit with 
$\chi^2_{\rm dof}  \simeq 1.0$ for the $[{\bf 70},\ell^+]$ multiplets.}
{\label{operators}
\renewcommand{\arraystretch}{1.5} % enlarge line spacing
\begin{tabular}{llrrl}
\hline
Operator & \multicolumn{4}{c}{Fitted coef. (MeV)}\\
\hline
$O_1 = N_c \ \1 $                                   & \ \ \ $c_1 =  $  & 556 & $\pm$ & 11       \\
$O_2 = \ell_q^i s^i$                                & \ \ \ $c_2 =  $  & -43 & $\pm$ & 47    \\
$O_3 = \frac{3}{N_c}\ell^{(2)ij}_{q}g^{ia}G_c^{ja}$ & \ \ \ $c_3 =  $  & -85 & $\pm$ & 72  \\
$O_4 = \frac{4}{N_c+1} \ell^i t^a G_c^{ia}$         & \ \ \            &     &       &     \\
$O_5 = \frac{1}{N_c}(S_c^iS_c^i+s^iS_c^i)$          & \ \ \ $c_5 =  $  & 253 & $\pm$ & 57  \\
$O_6 = \frac{1}{N_c}t^aT_c^a$                       & \ \ \ $c_6 =  $  & -25 & $\pm$ & 86    \\ 
\hline
$B_1 = t^8-\frac{1}{2\sqrt{3}N_c}O_1$               & \ \ \ $d_1 =  $  & 365 & $\pm$ & 169 \\
$B_2 = T_c^8-\frac{N_c-1}{2\sqrt{3}N_c}O_1$         & \ \ \ $d_2 =  $  &-293 & $\pm$ & 54 \\
$B_4 = 3 \ell^i_q g^{i8}- \frac{\sqrt{3}}{2}O_2$    & \ \ \            &     &       & \vspace{0.2cm}\\
\hline
\end{tabular}}
\end{table}

\section{Results}
Table \ref{operators} shows the values of the coefficients $c_i$ and $d_i$ obtained from a fit to available data.
Details concerning  the data used in the fit are presented elsewhere\cite{MS3}. One can see that the first order operator $O_1$ and the spin-spin operator $O_5$ are the most dominant ones, \emph{i.e.} $c_1$ and $c_5$ are large. The spin-orbit coefficient is negative, at variance with  previous studies\cite{MS2} but remains small in absolute value. The coefficient $c_3$ is twice smaller in absolute value as compared to that of Ref. [\refcite{MS2}]. We had to exclude the operator $O_4$ from the fit because it considerably deteriorated the fit. 

The SU(3)-flavor breaking operators play an important dynamical role as it can be seen from the values of the  coefficients $d_1$ and $d_2$. As all the matrix elements of $B_4$ cancel out for the available resonances,  it was not possible for us to obtain an estimation of $d_4$. 

\begin{figure}[h!]
\centerline{\includegraphics[width=9cm,keepaspectratio]{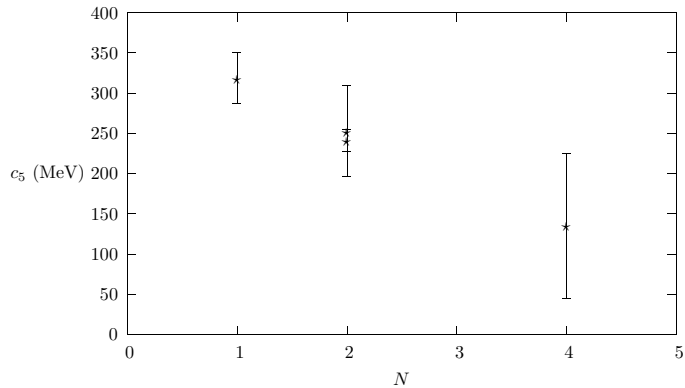}}
\caption{Evolution of the coefficient $c_5$ with the excitation energy corresponding to $N=1, 2$ and 4 bands in a harmonic oscillator notation.}
\label{c2}
\end{figure}

Figure  \ref{c2} shows the evolution of the spin-spin dynamical coefficient $c_5$ with the excitation energy.  Here we collected the presently known values with error bars for the orbitally excited states studied so far in the large $N_c$ expansion: $N=1$, Ref. [\refcite{SGS}], $N=2$ (lower value\cite{GSS03}, upper value\cite{MS3}) and $N=4$, Ref. [\refcite{MS1}]. This figure suggests that at large excitations, the spin-spin contribution vanishes.

%\end{center}

\section{Conclusions}
Here we have extended our previous workæ\cite{MS2} from SU(2) to SU(3). The present results confirm the dependence of the coefficients $c_1$, $c_2$ and $c_5$ as a function of excitation energy, namely that the contributions of the spin-dependent terms decrease with energy and eventually vanish at very large excitations. The analysis of $[{\bf 70},\ell^+]$ remains open.
More and better experimental data are needed to clarify the role of various terms contributing to the mass operator of the $[{\bf 70},\ell^+]$ multiplet.

\section*{Acknowledgments}
The work of one of us (N. M.) was supported by the Institut Interuniversitaire des Sciences Nucl\'eraires (Belgium).

\end{document}